\documentclass[a4paper,12pt,%
 preprint,
 amsmath,amssymb,aps,prl,showkeys,
floatfix,
]{revtex4-1}
\usepackage{color}
\usepackage{graphicx}
\usepackage{dcolumn}\usepackage{bm}
\usepackage{hyperref}

\def\figref#1{Fig.~\ref{#1}}
\def\eqnref#1{Eqn.~(\ref{#1})}

\begin{document}

\title{
Influence of irreversible demagnetization on the polarization of thermal radiation emitted by a hot cobalt wire}

\author{Armando Francesco Borghesani}\altaffiliation[Also: ]{CNISM Unit, Physics and Astronomy Dept., University of Padua}

\email{armandofrancesco.borghesani@unipd.it}
\author{Marco Guarise}\altaffiliation[Present address:]{ Dipartimento di Fisica e Scienze della Terra, University of Ferrara,  and Istituto Nazionale di Fisica Nucleare, Sezione di Ferrara, Ferrara, Italy}
\affiliation{Department of Physics and Astronomy\\
 University of Padua, Padua, Italy
}
\author{Giovanni Carugno}
 \affiliation{Istituto Nazionale di Fisica Nucleare, Sezione di Padova\\
 Padua, Italy
}
\begin{abstract}
We report measurements of the polarization \(P\) of 
thermal radiation emitted by a 
cobalt  wire in the temperature 
range from \(T\approx 400\,\)K up to melting. The radiation is linearly polarized perpendicular to the wire.   
 \(P\) decreases from \(30\,\%\) just above room temperature down to \(6.5\,\%\) near melting and does not show any particular behavior neither at the martensitic {\em hcp}$-${\em fcc} 
transition at \(\approx 700\,\)K nor at the Curie point at \(
\approx 1400\,\)K. However, 
\(P\) shows a rapid decrease for \(T\gtrsim 1000\,\)K and, contrary to previous 
measurements with tungsten wires,
it hysteretically behaves if the temperature 
 change is reversed. This behavior is rationalized by 
 accounting for the irreversible 
 thermal demagnetization  of the wire  with magnetic domain size change.
\end{abstract}

\maketitle

The dynamics of the domain structure of magnetic materials has long attracted a great deal of interest.
The  domain wall (DW) motion is thoroughly investigated for its relevance in fundamental science and applications. Thermally activated~\cite{nattermann2001}, magnetic field~\cite{metaxas2007} and/or current~\cite{duttagupta2017} driven DW creep and flow, DW depinning~\cite{zapperi1998,chauve2000} or precessional modes~\cite{hayashi2007}, thermally driven diffusive DW motion~\cite{leliaert2015}, thermally- or field driven domain reversal~\cite{ounadjela1997,brands2006,dmitriev2016}, magnetic viscosity~\cite{gaunt1977,gaunt1986},
 among many other topics, are studied with a number of techniques in several samples of different composition, size, and geometry because of their relevance in many applications, including geophysics and paleomagnetism~\cite{kelso1988,heider1988}.

The coupling of radiation with the sample properties is exploited to investigate its domain structure and dynamics via the magnetooptical Kerr effect (MOKE)~\cite{fowler1954,weller1994,ono1995}. As it is based on the rotation of the polarization plane and intensity change of visible light reflected off 
a magnetic material, MOKE is limited to provide pieces of information on the material surface. For instance, the surface domain structure, magnetization switching and reversal in amorphous microwires under different experimental conditions have been succesfully studied by MOKE techniques~\cite{chizhik2001,chizhik2002,chizhik2004,chizhik2010}.

However, pieces of information on the bulk properties and structure of a sample can also be gathered  by investigating the properties of the thermal radiation emitted by a thin metallic wire.  Actually, a hot body at temperature \(T\) and of size larger than the typical thermal wavelength \(\lambda_{T}= hc/k_\mathrm{B}T\), where \(h\) is the Planck's constant, \(c\) is the speed of light, and \(k_\mathrm{B}\) is the Boltzmann constant, emits incoherent and unpolarized radiation. Nonetheless, if the phase space available for the collective fluctuations of the electron gas is reduced by geometrically limiting the radiator size, the thermal radiation acquires a degree of linear polarization~\cite{agdur1963}. 
This fact is important because in recent years there is an 
interest to produce nano-heaters and -light sources for applications in applied physics and engineering~\cite{ingvarsson2007,au2008,klein2009}. 
Recently, we  have 
carried out measurements of the linear polarization \(P\) of the radiation emitted by hot 
tungsten wires. We have 
found that it is partially polarized perpendicular to the symmetry axis of the wire because the thermally driven collective transverse fluctuations of the electron sea are limited by the wire boundaries and that its polarization degree decreases from \(P\approx 30\,\%\) for \(T\approx 500\,\)K down to \(\approx 15\,\%\) just before melting at \(T\approx 3700\,\)K. We have shown that the experimental behavior of \(P\) is reproduced by computing the absorption efficiency of radiation impinging on a cylindrical object of known radius and that it is intimately related to the bulk optical properties of the material~\cite{Borghesani2016}.
Therefore, we have decided to investigate the polarization of the thermal radiation emitted by a thin cobalt wire in order to see if pieces of information on the magnetic domain structure and thermally driven dynamics of the material can be obtained from optical measurements on a macroscopic sample.

Cobalt is primarily chosen because of its interesting magnetic and crystalline properties. Its ferromagnetic-paramagnetic transition occurs at a Curie temperature of \(\approx 1400\,\)K.
Moreover, it shows a  martensitic transition at \(T\approx 700\,\)K from the low-\(T\) {\em hcp}- to the high-\(T\) {\em fcc} structure. At low temperature the hexagonal axis is the direction of easiest magnetization, whereas in the {\em fcc} phase at high temperature the metal becomes isotropic~\cite{nishizawa1983}. 
As the present experiment is carried out as a function of 
\(T\) up to melting, cobalt offers a unique opportunity to investigate a rich realm of behaviors. 

In this Letter we show that a possible explanation of the experimental outcome can be traced back to a progressive irreversible sample demagnetization with a decrease of the transverse size of the magnetic domains as the wire temperature is increased. 

Apparatus and technique are the same used for investigating the polarization of the thermal radiation emitted 
by tungsten wires and are thoroughly described in literature~\cite{Borghesani2016}. We recall here the main features of the experiment. A  $7\,$mm long cobalt wire of radius \(R_{0}=50\,\mu\)m  (\(99.99\,\% +\) purity, Goodfellow Cambridge Ltd) is mounted in a nonmagnetic vacuum cell. The wire is heated by an adjustable d.c. current that sets its temperature, 
which is linear in the d.c. 
dissipation. The uncertainty on \(T\) is \(\approx \pm 10\,\)K~\cite{Borghesani2016}. A weak low-frequency (\(f\approx 2\,\)Hz) a.c. modulation is superimposed on the d.c. current 
to allow the use of lock-in (LI) detection techniques.
Two ZnSe lenses image the wire  onto a a liquid N\(_{2}\) cooled photovoltaic HgCdTe detector (Fermionics, mod. PV-12-0.5) of spectral range \(0.5\,\mu\mbox{m}\le\lambda\lesssim 12\,\mu\mbox{m}\).  The detector output feeds an amplification stage composed by a transimpedance amplifier (Fermionics, PVA-500-10),  a linear amplifier (EG\&G, PARC, mod. 113), and a LI amplifier (Stanford Research Systems, mod. SR830) and is digitized and recorded by a P.C. The thermal radiation is analyzed by a ZnSe wire grid, infrared polarizer (Thorlabs, WP25H-Z) mounted on a rotary goniometer. As \(f\) is very low, 
the LI output is averaged for well over \(60\,\)s for every position of the goniometer.

\begin{figure}[b!]
\includegraphics[width=\columnwidth]{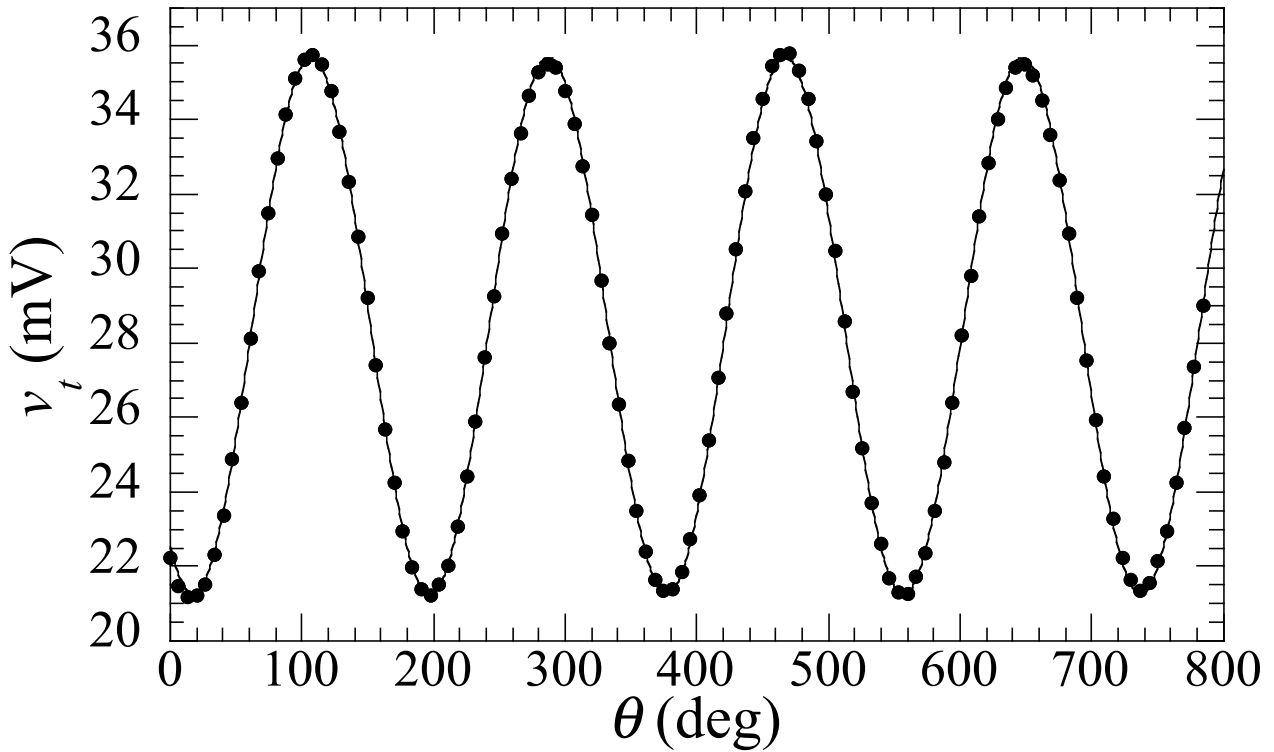
}\caption{\small \(v_{t}\) vs \(\theta\). The error bars are of the size of the points.\label{fig:malus}}
\end{figure}

The LI output follows the Malus law 
\(v_{t}= v_{u}+ v_{p}\cos^{2}{\left( \theta -\theta_{0}\right)}\).
\(v_{u}\) and \(v_{p}\) are the contributions of the unpolarized and polarized radiation components, respectively. \(\theta\) is the polarizer angle and \(\theta_{0}\) is an unessential initial angle. 
A typical detector output \(v_{t}\) as a function \(\theta\) is shown in \figref{fig:malus}.  

The average polarization degree is computed as the polarization contrast
\begin{equation}
P= \frac{v_{p}}{2v_{u}+v_{p}}
\label{eq:pol}\end{equation}

\begin{figure}[b!]
\includegraphics[width=\columnwidth]{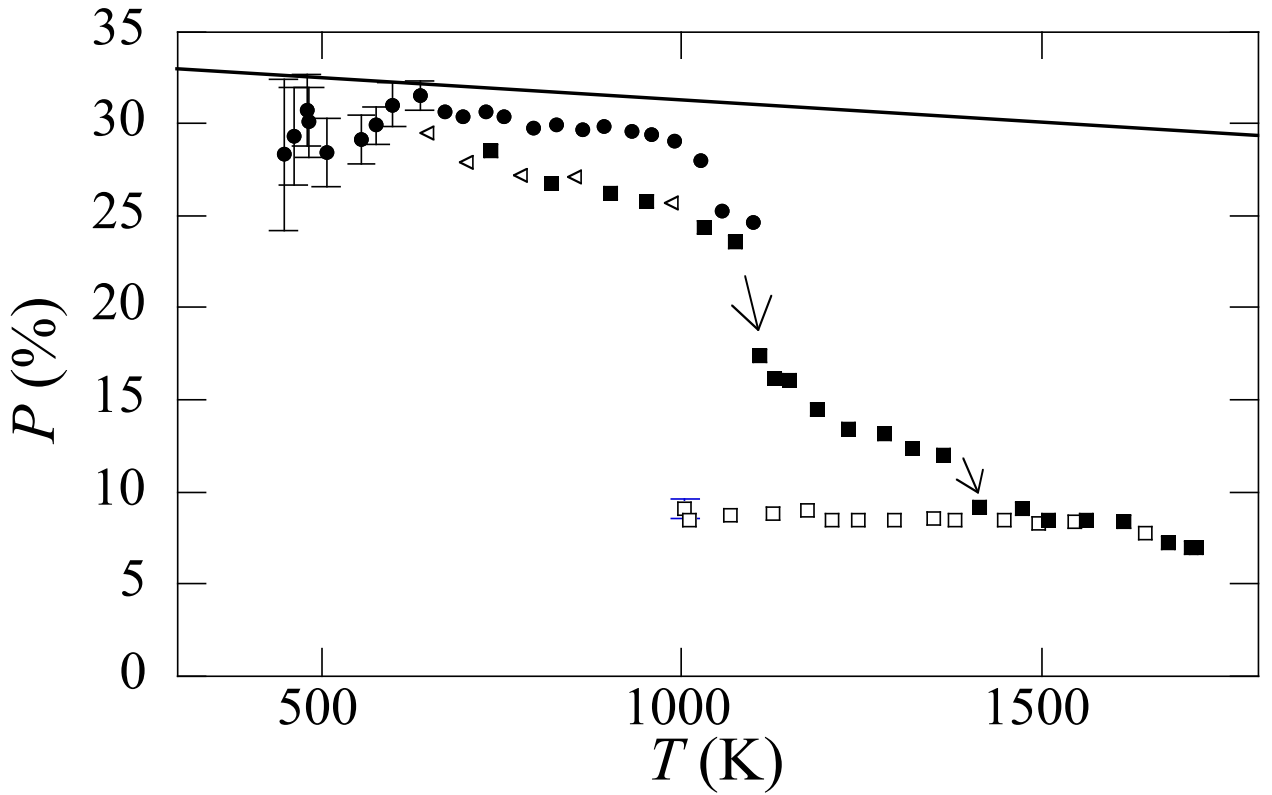}\caption{\small \(P\) vs \(T\) for two cobalt wires.  Line: theory. 
For the meaning of the symbols, see text.\label{fig:isteria}}
\end{figure} 

In \figref{fig:isteria} \(P\) is shown as a function of \(T\) for two 
 wires. At variance with the tungsten case~\cite{Borghesani2016} in which \(P\) monotonically decreases with increasing \(T\), the cobalt wires show a hysteretic behavior. \(T\) is increased at first starting at \(T\approx 400\,\)K up to \(T\approx 1100\,\)K (closed circles). In this  
range \(P\) decreases from  the theoretically predicted value \(P\approx 33\,\%\) down to \(P\approx 25\,\%\).
From \(T\approx 1100\,\)K, \(T\) was progressively decreased back to \(T\approx 650\,\)K (triangles) with \(P\) not retracing the values measured during the heating phase.  Then, \(T\) was again increased (closed squares) up to melting. During reheating up to \(T\lesssim 1100\,\)K, \(P\) does not show any significant hysteresis. At about \(T\approx 1100\,\)K, there apparently is a sudden decrease of \(P\) with increasing \(T\). We note, however, that, whereas nearby data points were recorded at an interval of \(\approx 12\,\)hrs from each other, the time interval between the data connected by arrows was 2 days. Although the wire, owing to its small mass and thermal capacity reaches thermal equilibrium in a few minutes after \(T\) is changed, it appears that its internal structure is following a very slow dynamics.
This point of view is confirmed by the behavior of \(P\) observed in another similar wire (open squares in \figref{fig:isteria}), which was first kept at high temperature (\(T\approx 1600\,\)K) for nearly one month in order to calibrate the optical detection system before the polarization measurements were carried out. 
For this long annealed wire, when \(T\) was decreased,  \(P\) remained constant at the value reached by the first wire close to melting.
Additionally, we observe that \(P\) does neither show any particular behavior at the {\em hcp-fcc} martensitic transition at \(T\approx 700\,\)K~\cite{prem2001} nor at the Curie temperature \(T_{c}\approx 1400\,\)K.

The polarization of a wire of homogeneous material can be predicted by computing the absorption efficiency factors \(Q_\mathrm{abs}^{\parallel,\,\perp}(\lambda,T,R)\) for transverse electrical (TE, \(\perp\)) and transverse magnetic (TM, \(\parallel\)) modes of the 
radiation field impinging on a indefinitely long cylinder of radius \(R\) ~\cite{agdur1963,ruoso2009,kardar2012,Borghesani2016} provided that the dependence of the relative dielectric constant \(\varepsilon_{r}\) on 
\(T\)
and \(\lambda\) 
is known. For cobalt \(\varepsilon_{r}\) is given by a Drude-type form whose parameters are given in literature~\cite{makino1982,ordal1983,ordal1985}.
The observed polarization is given as a function of \(T\) and \(R\) by
\begin{equation}
P(T,R) = \frac{\langle Q_\mathrm{abs}^{\perp}\rangle-\langle Q_\mathrm{abs}^{\parallel}\rangle}{\langle Q_\mathrm{abs}^{\perp}\rangle+\langle Q_\mathrm{abs}^{\parallel}\rangle}
\label{eq:ptr}
\end{equation}
in which the average is taken over the accessible  wavelength spectrum
\begin{equation}
\langle Q_\mathrm{abs}^{\dagger} \rangle = \frac{1}{F}\int \mathcal{D}(\lambda)B(\lambda,T) 
Q_\mathrm{abs}^{\dagger}(\lambda,T, R)
\,\mathrm{d}\lambda \quad (\dagger = \perp, \parallel)
\label{eq:averages}\end{equation} 
\(B(\lambda,T)\) is the Planck's distribution, \(\mathcal{D}(\lambda)\) is the detector responsivity,
 and the normalization is \(F=\int \mathcal{D}(\lambda)B(\lambda,T) \,\mathrm{d}\lambda \). 
The absorption efficiencies for the two modes \((\dagger = \perp, \parallel)\) are given by
\begin{equation}
Q_\mathrm{abs}^{\dagger}(\lambda,T,R)= \frac{2}{k R} \left[ \mathtt{Re} \left(a_{0}^{\dagger}+2\sum\limits_{m=1}^{\infty}a_{m}^{\dagger}\right)\right.
+\left.\left(\vert a_{0}^{\dagger}\vert^{2}+2\sum\limits_{m=1}^{\infty}\vert a_{m}^{\dagger}\vert^{2}\right) \right] 
\label{eq:qabs}\end{equation}
with \(k=2\pi/\lambda\).
The coefficients \(a_{m}^{\dagger}\) are obtained by enforcing the boundary conditions as
\begin{eqnarray}
a_{m}^{\perp}&=&\frac{J^{\prime}_{m}(nkR) J_{m}(kR) -nJ_{m}(nkR)J_{m}^{\prime}(kR)}{J_{m}^{\prime}(nkR) H_{m}^{(2)}(kR) -nJ_{m}(nkR) H_{m}^{(2)\prime}(kR)}
\label{eq:aperp}\\
a_{m}^{\parallel}&=&\frac{nJ_{m}^{\prime}(nkR)J_{m}(kR)-J_{m}(nkR)J_{m}^{\prime}(kR)}{nJ_{m}^{\prime}(nkR) H_{m}^{(2)}(kR) -J_{m}(nkR) H_{m}^{(2)\prime}(kR)}
\end{eqnarray}
in which \(n=\sqrt{\varepsilon_{r}}\) is the complex refraction index, \(J_{m}\) and \(H_{m}^{(2)}\) are the Bessel functions of first kind and the Hankel functions of second kind, respectively.
The solid line in \figref{fig:isteria} is the prediction of \eqnref{eq:ptr} for a wire of nominal radius \(R_{0}=50\,\mu\)m. Evidently, the prediction of \(P\) for a homogeneous material completely disagrees with the experimental data, except at the lowest \(T\) where \(P\approx 30\,\%\) is a universal limit for \(\lambda_{T}\ge R\), independent of the material~\cite{agdur1963}. 
For any other \(T\), \(P\) is lower than predicted.

\begin{figure}[t!]
\includegraphics[width=\columnwidth]{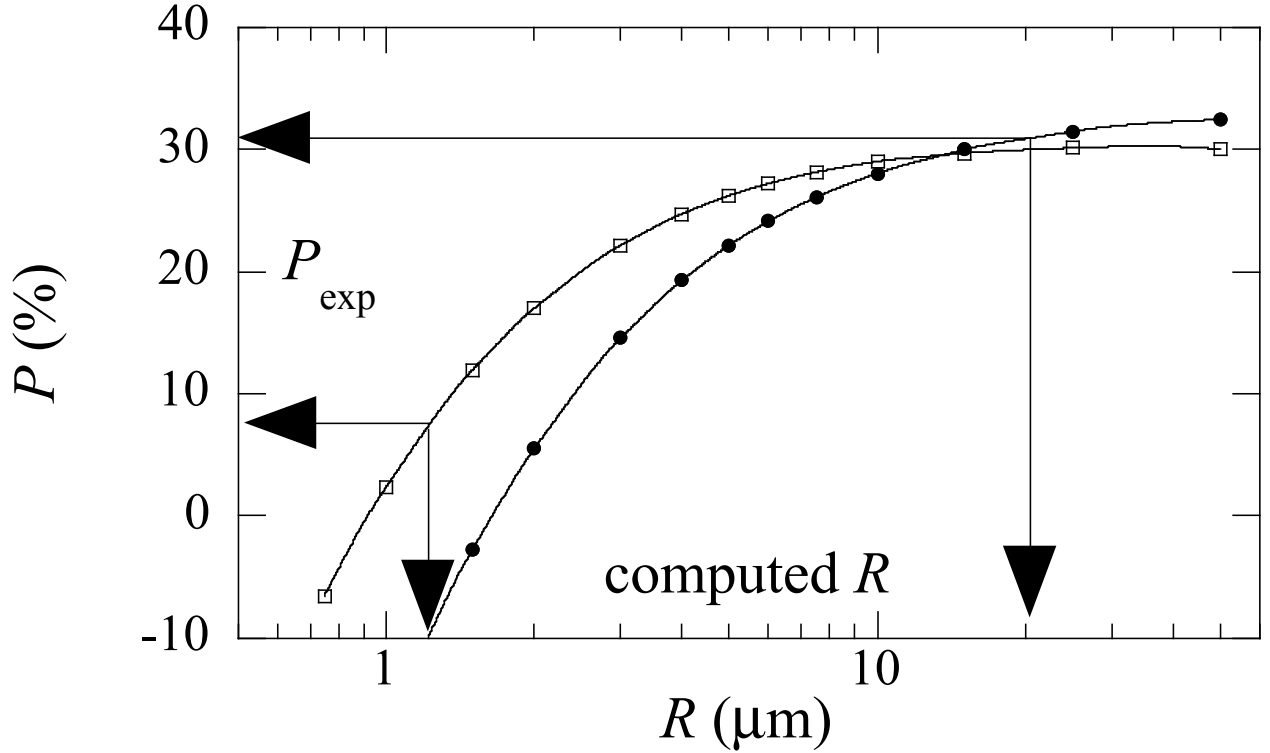}\caption{\small 
\(P\) vs \(R\) computed for \(T=500\,\)K (closed points) and for \(T=1500\,\)K (open squares). For \(P<0\) the polarization is parallel to the wire axis. The lines are an eyeguide only.\label{fig:PofR800K}}
\end{figure}

However, the computed \(P\) strongly depends on \(R\) at any \(T\) as shown in \figref{fig:PofR800K}. As \(R\) 
decreases, \(P\) decreases as well, and, for very small \(R\), it also becomes negative, i.e., parallel to the wire  axis.
Upon decreasing \(R\) the phase space available to the transverse modes of  collective charge fluctuations is shrinked and \(P\) 
gets smaller than in larger wires.  At the same time, the intensity radiated per unit wire surface \(j\) is practically independent of \(R\) and the radiated intensity  
\(I\propto 2v_{u}+v_{p}\), which is the other experimentally measured quantity, is proportional to \(R\), \(I = jR\). As a consequence, a bundle of \(N=R_{0}/R\) wires of radius \(R<R_{0}\) would radiate the same \(I\) as the larger wire of radius \(R_{0}\) but \(P\) would be smaller.

\begin{figure}[t!]
\includegraphics[width=\columnwidth]{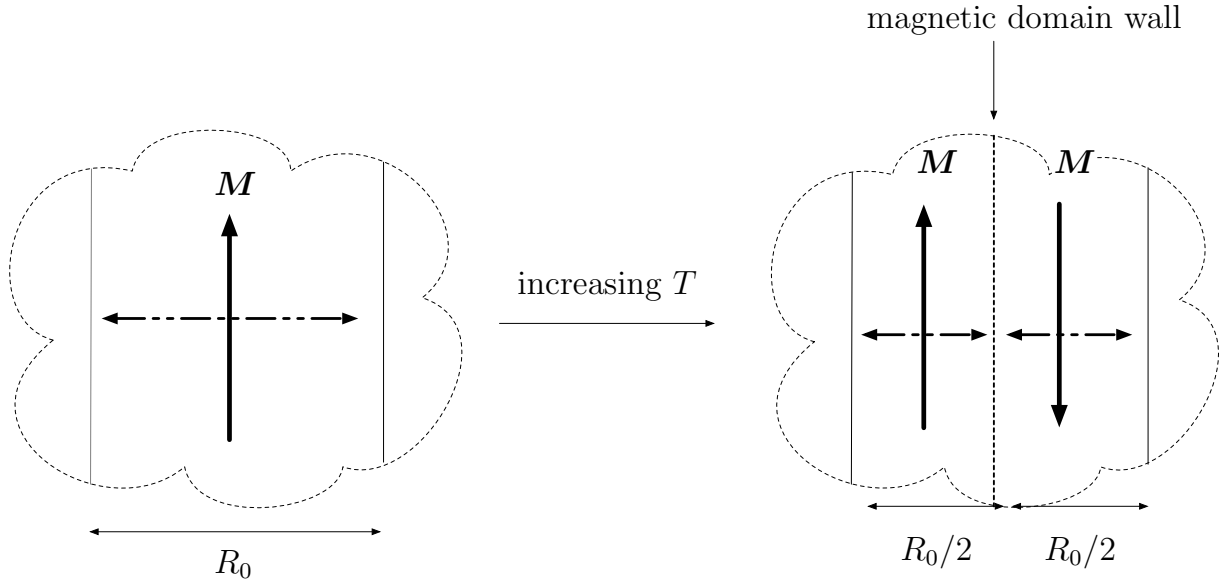}\caption{\small Crude model to rationalize the decrease of \(P\) with increasing \(T\) due to thermally induced wire demagnetization. \(M\): remanent magnetization. Dash-dotted lines:  range available for transverse modes of collective charge fluctuations.
\label{fig:demag}}
\end{figure}

This observation suggests that thermally induced demagnetization of the cobalt wire might explain the observed behavior of \(P\) as a function of \(T\). Let us consider the following very crude model depicted in \figref{fig:demag}. Let us assume that the wire at room temperature consists of a dominant magnetic domain with the magnetization aligned parallel to the wire long axis because of magnetoelastic anisotropy resulting from the coupling between the internal stresses due to the drawing production process and magnetostriction. 
The transverse modes of charge fluctuations can span the whole wire diameter (left part of \figref{fig:demag}). Upon increasing \(T\), thermally activated DW motion takes places and a domain with reversed magnetization grows larger in order to reduce the magnetic energy (right part of \figref{fig:demag}). As the typical thickness of the DW is much larger than the typical electron wavelength at the Fermi level~\cite{tatara2008}, the DW offers a enhanced resistance~\cite{cabrera1974a,cabrera1974b,rudiger1999,ebels2000,kent2001,leven2005,boonruesi2019} across the wire. 
This impedance mismatch at the DW would   produce reflection of the collective charge fluctuations transverse modes thereby reducing their available spatial range and long wavelength cutoff. In such a way \(P\) is reduced but the radiated intensity  
would remain the same. By increasing \(T\), this process keeps occurring, thereby leading to a further subdivision of the wire in gradually smaller magnetic domains.

We believe that the DW motion is thermally driven diffusion~\cite{leliaert2015} for several reasons. 
Current-driven DW motion is ruled out because it requires current densities at least in excess of \(\approx 10^{9}-10^{10}\,\)A/m\(^{2}\)~\cite{duttagupta2017,diazpardo2019} whereas in our experiment the maximum current density is \(\lesssim1.8\times 10^{8}\,\)A/m\(^{2}\).  
Moreover, in the present case the current is flowing parallel rather than perpendicular to the DW's. 
Magnetic field-driven DW motion is also ruled out because the wire is mounted in a magnetic material free environment. The only magnetic field in the experiment is generated by the current itself flowing in the wire, lies in planes perpendicular to the wire axis, and does not exceed the value \(H\approx 4\,\)kA/m at the wire circumference for the highest current used in the experiment. Moreover, as the typical time scale of the observed dynamics in our experiment is \(\tau\sim 10^{5}\,\)s, the estimate of the strength \(E\) of activation energy barriers for thermal activation of magnetization reversal would yield \(E/k_\mathrm{B}T> 35\)~\cite{gaunt1986}, a value which seems quite too large. 
Additionally, we note that diffusive DW motion may show a hysteretic behavior~\cite{nattermann2001} as observed in our experiment.
 
\begin{figure}[b!]
\includegraphics[width=\columnwidth]{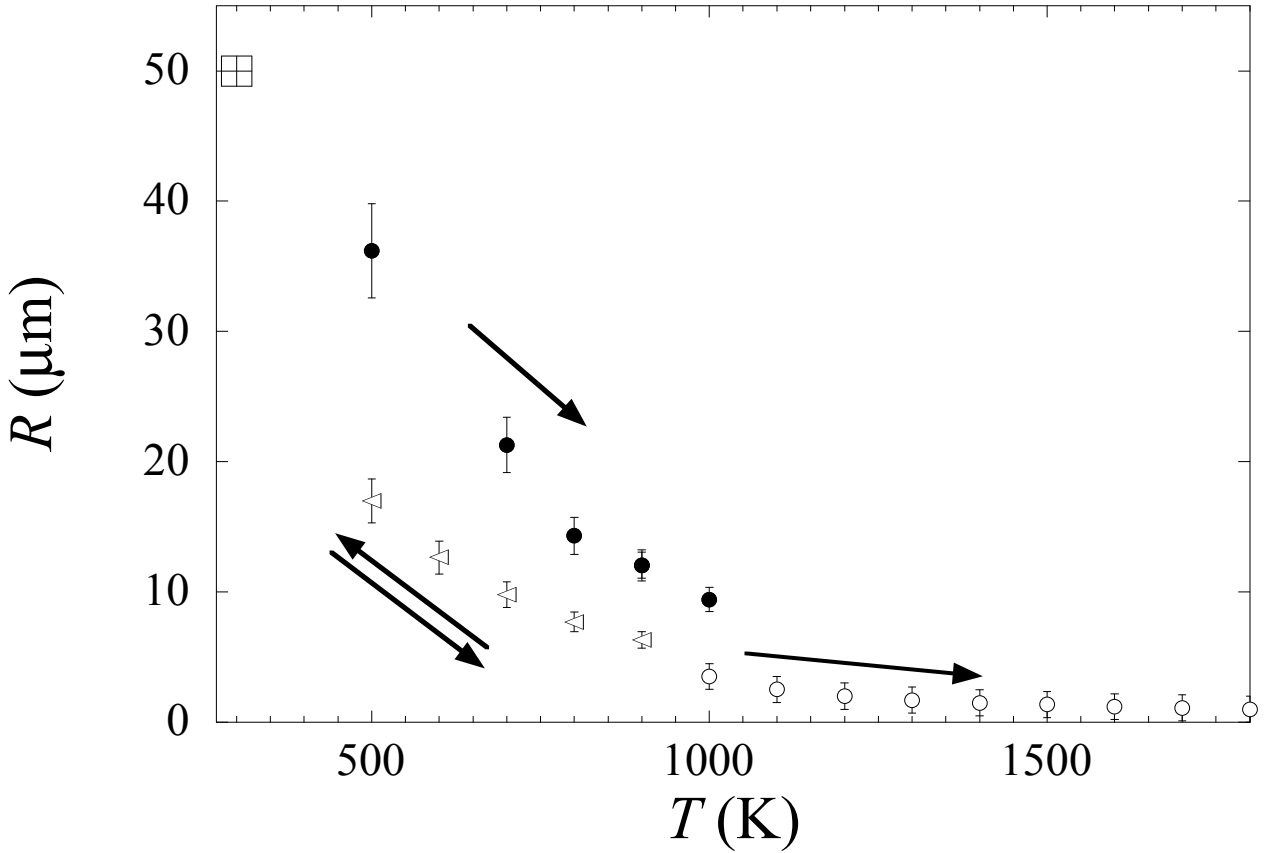}\caption{\small \(R(P)\) vs \(T\). 
Crossed square: nominal 
radius. Closed circles: initial heating. Triangles:  cooling and reheating. Open circles:  final heating. Arrows:  direction of \(T\) changes. \label{fig:RdaP}}
\end{figure}

According to the present model, 
we can estimate 
the radius \(R(T)\) of the wires in the bundle 
by solving for \(R\) the equation \(P(T,R(T))= P_\mathrm{exp}\), where \(P_\mathrm{exp}\) is the measured polarization value, as shown in \figref{fig:PofR800K}. The resulting \(R\) is shown as a function of \(T\) in \figref{fig:RdaP}. Within the conceptual framework of this model, \(R\) can be thought of as an estimate of the average transverse size of the magnetic domains. Upon increasing \(T\) from room temperature, \(R\) decreases from the nominal value of \(50\,\mu\)m down to \(\approx 10\,\mu\)m. After the polarization drop for \(T\approx 1100\,\)K, during the cooling phase, \(R\) shows hysteresis by remaining smaller than during the initial heating. Upon the final reheating, \(R\) steadily decreases and reaches the value \(R\approx 1 \,\mu\)m just before melting.
This behavior is coherent with a thermally induced demagnetization process of the sample as detected in several different types of measurements~\cite{kelso1988,heider1988,ono1995,szmaja1995}.
\begin{figure}[b!]
\includegraphics[width=\columnwidth]{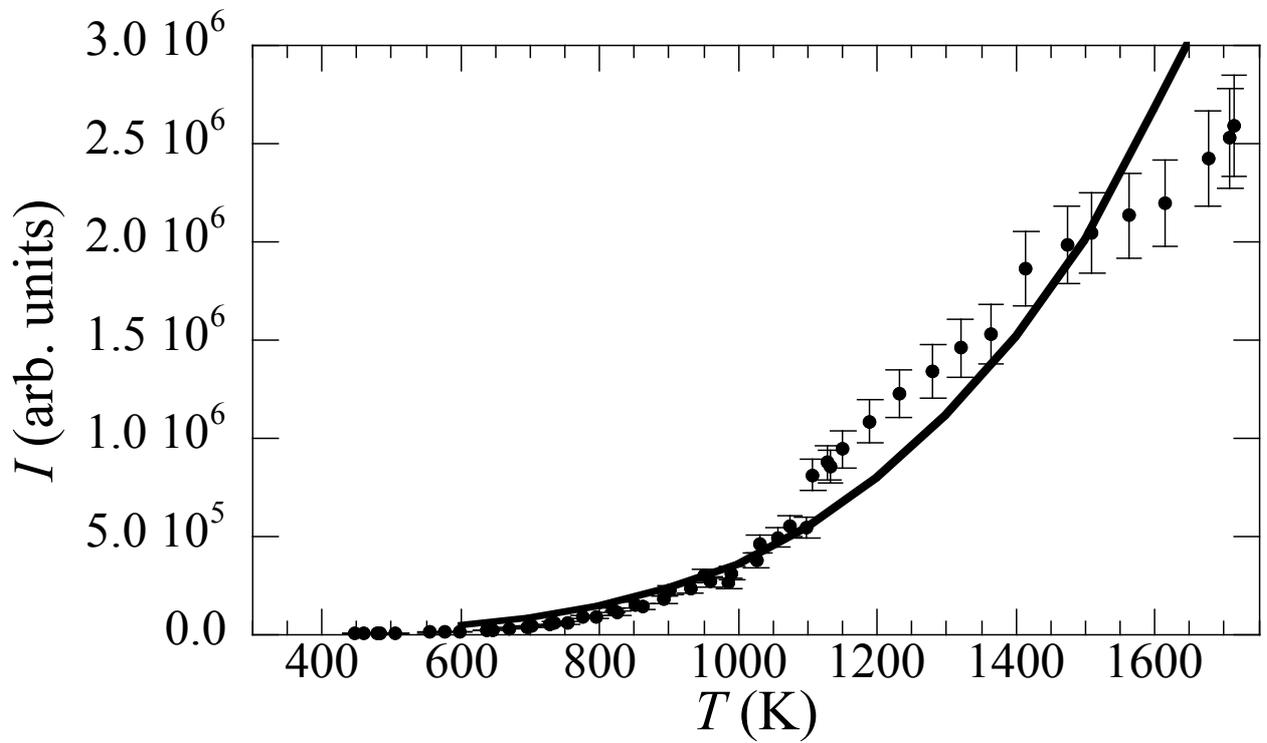
}\caption{\small \(I\) vs \(T\). Points: experiment. Solid line: theory.
 \label{fig:IvsRdaP.eps}}
\end{figure}
At the same time, according to our expectations, the total radiated intensity \(I\) should be insensitive to the bundle structure and should only depend on \(T\). Actually, this is the case, as shown in \figref{fig:IvsRdaP.eps}, where the measured \(I\) is compared with the theoretical prediction \(I\propto \langle Q^{\perp}_\mathrm{abs}\rangle+ \langle Q^{\parallel}_\mathrm{abs}\rangle\) evaluated at the temperature dependent \(R\) of 
\figref{fig:RdaP}. The  agreement between experiment and theory is quite satisfactory.

In conclusion, we can state that optical measurements on macroscopic magnetic materials can shed light on the temperature evolution of the magnetic domain structure of the material itself. Actually, the piece of information that this kind of measurements can provide is only macroscopic and, in some sense, of thermodynamic nature because it does not 
give any insight into the microscopic structure of the sample. At the same time, we would like to emphasize that the present results are very satisfactory even taking into account the extreme sensitivity of the magnetic material properties to, among others,  composition, manufacturing, annealing, and ageing of the sample.

\begin{acknowledgments}
The authors gratefully acknowledge the technical assistance of E. Berto and G.P. Galet.
\end{acknowledgments}

\end{document}